\documentstyle[amsmath,amssymb,11pt,epsfig]{article}
\begin{document}
\title{\bf{Searching for new physics in Upsilon decays}
\thanks{Research under grant FPA2002-00612.}}
\author{
Miguel Angel Sanchis-Lozano$^{a,b}$\thanks{Email: Miguel.Angel.Sanchis@uv.es}
\vspace{0.4cm}\\
(a) Instituto de F\'{\i}sica
Corpuscular (IFIC), Centro Mixto Universidad de Valencia-CSIC \\
(b) Departamento de F\'{\i}sica Te\'orica, Universidad de Valencia \\
Dr. Moliner 50, E-46100 Burjassot, Valencia (Spain)}
\date{}
\maketitle
\begin{abstract} 

	We examine some possible experimental consequences of 
        new physics on the spectrum and decays of bottomonium 
        states below $B\bar{B}$ threshold. In addition to lepton 
        universality breaking in Upsilon decays, large widths 
        of pseudoscalar $\eta_b$ resonances 
        and mixing with a non-standard CP-odd light Higgs boson
        might smooth and shift the signal peak from hindered radiative 
        M1 transitions between $\Upsilon$ and $\eta_b$ states  
        in the photon energy spectrum, as searched by CLEO. 
        We also stress the relevance of forthcoming results from CLEO 
        on leptonic branching fractions of $\Upsilon$
        resonances to definitely check our hypothesis.

\end{abstract}
\vspace{-11.5cm}
\large{
\begin{flushright}
  IFIC/04-1\\
  FTUV-04-0107\\
  January 7, 2004\\
  hep-ph/0401031
\end{flushright} }
\vspace{9.cm}
\begin{small}
PACS numbers: 14.80.Cp, 13.25.Gv, 14.80.-j \\
Keywords: New physics, non-standard Higgs, bottomonium decays, 
lepton universality
\end{small}
 
\vskip 1.cm
 
The possibility of new physics showing up in leptonic decays
of bottomonium was examined in \cite{mas03,mas02}. There
we advocated  the existence of a non-standard (CP-odd) 
light Higgs boson (denoted as $A^0$)
mediating the annihilation of the Upsilon vector resonance
into a lepton pair through an intermediate 
$\eta_b^*$ state subsequent to a magnetic dipole (M1) transition
yielding a soft (unobserved) final-state photon.
Then an $\lq\lq$apparent'' breaking of lepton universality
\footnote{In the sense that once the
Higgs contribution were taken into account, lepton universality
would be restored} would appear since
the Higgs-mediated contribution would be unwittingly ascribed to the 
leptonic decay channel in the experimental measurement 
(notably in the tauonic mode because of the
employed missing-energy technique) thereby inducing a dependence 
of the branching fraction (BF) on the leptonic species.

Hitherto, we have assumed in our theoretical analysis that
the $A^0$ mass was close to the 
bottomonium resonances below $B\bar{B}$ threshold; thus
the Higgs propagator in the diagram of Fig.1b should enhance the 
decay rate ultimately implying moderate values
of the parameter $\tan{\beta}$ - defined as the ratio of
the Higgs vacuum expectation values in a two Higgs doublet
model (2HDM) \cite{gunion} - to account for the postulated 
new physics effect in $\Upsilon$ leptonic decays. 
In this Letter we relax this condition to some extent, moreover
exploring the experimental consequences of our conjecture on bottomonium
spectroscopy and hindered radiative
transitions from $\Upsilon$ resonances in view of the current
failure to detect the 
$\eta_b(1S)$ and $\eta_b(2S)$ states \cite{cleo02,cleo03}.
Admittedly, the lack of such experimental observation in itself can be
hardly used as an argument in favor of new physics, but nonetheless one 
should keep an open mind in
connection with the ideas developed in this work.

\begin{figure}
\epsfysize=5.5cm
\centerline{\hbox{{\hfil\epsfbox{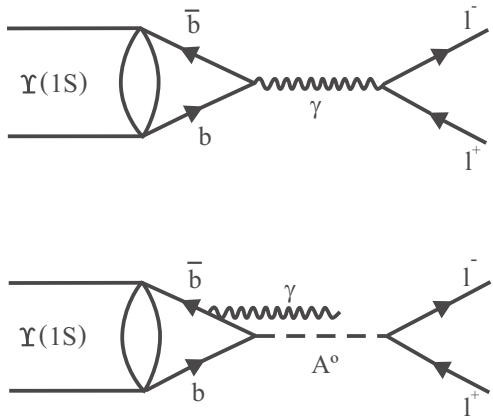}\hfil}}}
\caption{(a)[upper panel]: Electromagnetic 
annihilation of a $\Upsilon(1S)$ resonance into a charged
lepton pair through a virtual photon;
(b)[lower panel]: Hypothetical annihilation of an  
intermediate $\eta_b^*$ state (subsequent to a M1 structural transition
yielding a soft photon)  
into a charged lepton pair through a CP-odd Higgs particle 
denoted by $A^0$.}
\end{figure}

In order to assess the relative importance of the postulated new physics
contribution, we defined in \cite{mas03,mas02} the BF's ratio: 
\begin{equation}
{\cal R}_{\ell} = \frac{{\cal B}_{\Upsilon{\rightarrow}\gamma_s\ell\ell}}
{{\cal B}_{\Upsilon \to \ell\ell}}\ \ \ \ ;\ \ \ \ \ell=e,\mu,\tau
\label{eq:ratio}
\end{equation}
where ${\cal B}_{\Upsilon{\rightarrow}\gamma_s\ell\ell}$ and
${\cal B}_{\Upsilon{\rightarrow}\ell\ell}$ refer to the
Higgs-mediated and standard electromagnetic 
contributions to the $\Upsilon$ leptonic decay,
respectively.
Focusing on the tauonic channel,
${\cal B}_{\Upsilon{\rightarrow}\gamma_s\tau\tau}$
was estimated as the $\lq\lq$excess''
${\cal B}_{\Upsilon{\rightarrow}\tau\tau}
-{\cal B}_{\Upsilon{\rightarrow}\ell\ell}$ with either $\ell=e,\ \mu$
from available experimental data \cite{pdg}; thereby
${\cal R}_{\tau}$ was found to be of order $10\%$ in 
both $\Upsilon(1S)$ and $\Upsilon(2S)$ decays
although with a considerable experimental uncertainty \cite{mas03,mas02}. 
Nevertheless, we performed a hypothesis test concluding 
that lepton universality (predicting ${\cal R}_{\tau}=0$)
could be rejected at a 10$\%$ level of significance.

Two theoretical approaches were used in \cite{mas03}
to deal with the $\Upsilon \to \gamma_s\ell\ell$ decay. Firstly, 
we applied time-ordered perturbation theory considering
a two-step process: a prior M1-transition yielding
a $b\bar{b}$ pseudoscalar virtual state
followed by its annihilation into a lepton pair mediated by the
$A^0$ (see Fig.1b).
Alternatively, we relied 
on the separation between 
long- and short-distance physics in accordance with the main lines of 
a non-relativistic effective theory like NRQCD 
\cite{bodwin}, albeit replacing a gluon 
by a photon in the usual Fock decomposition 
of hadronic bound states. Different results for the
final widths come out in each approach, as we shall see 
below (see \cite{mas03} for a lengthier discussion).
 
On the one hand, factorization \`a la NRQCD \cite{bodwin}
of the decay width leads to
\begin{equation}
\Gamma_{\,\Upsilon\to\gamma_s\,\ell\ell} =
{\cal P}^{\Upsilon}(\eta_b^*\gamma_s) 
\times {\Gamma}_{\eta_b^*\to\ell\ell}
\label{eq:factorization}
\end{equation}
where ${\cal P}^{\Upsilon}(\eta_b^*\gamma_s)$ denotes
the probability for a Fock component in the $\Upsilon(nS)$ 
($n=1,2,3$) resonance containing an almost on-shell $\eta_b^*$
state and a soft photon $\gamma_s$; in Refs.\cite{mas03,mas02} this
probability was estimated by means of the well-known formula
connecting $\Upsilon(nS)$ and $\eta_b(nS)$ states through a direct
M1-transition: 
\begin{equation}
{\cal P}^{\Upsilon}(\eta_b^*\gamma_s)\ \approx\ 
\frac{\Gamma^{M1}_{\Upsilon{\rightarrow}\gamma_s\eta_b}}
{\Gamma_{\Upsilon}}\ {\simeq}\ 
\frac{1}{\Gamma_{\Upsilon}}\ \frac{4\alpha Q_b^2}{3m_b^2}\ 
\Delta E_{hs}^3\ \sim 10^{-5}-10^{-3}
\label{eq:probability}
\end{equation}
where $\Gamma_{\Upsilon}$ stands for the resonance full width, 
$\alpha \simeq 1/137$ denotes the  
fine-structure constant,
$Q_b=1/3$ is the bottom-quark electric charge
and $m_b=M_\Upsilon/2 \simeq 5$ GeV; we took 
the hyperfine $\Upsilon-\eta_b$ mass splitting
$\Delta E_{hs}$ varying over the broad range
$35-150$ MeV.
Note that the intermediate $\eta_b^*$ 
state actually should not be too far off-shell 
because of the small photon energy given by $\Delta E_{hs}$.
So we will not distinguish between $\eta_b$ and $\eta_b^*$
hereafter; besides $\eta_b$ denotes collectively
any $\eta_b(nS)$ $(n=1,2,3)$ state though we have focused
on the $\eta_b(1S)$ because of currently more precise data
on the leptonic BF of the $\Upsilon(1S)$ \cite{pdg}.

Thus, the final decay width for the whole process can be re-expressed 
from Eq.(\ref{eq:factorization}) as
\begin{equation}
\Gamma_{\,\Upsilon\to\gamma_s\,\ell\ell}\ \simeq\ 
\frac{\Gamma^{\,M1}_{\,\Upsilon\to\gamma_s\eta_b}} 
{\Gamma_{\Upsilon}}\ \Gamma_{\eta_b\to\ell\ell}
\label{eq:finalwidth} 
\end{equation} 

By requiring the BF's ratio ${\cal R}_{\tau}$ in Eq.(\ref{eq:ratio}) 
to be of order $10\%$, we found in \cite{mas03}
that $\tan{\beta}$ in a 2HDM(II) has to stay over the range:
$7\ {\lesssim}\ \tan{\beta}\ {\lesssim}\ 21$, 
depending on the value of $\Delta E_{hs}$, namely from 150 MeV 
down to 35 MeV. Actually this $\tan{\beta}$ interval
corresponds to a particular choice of the Higgs mass value, 
in between the $\Upsilon(1S)$ and $\Upsilon(2S)$
masses. Here we relax somewhat this condition:
Fig.2 shows the $\tan{\beta}$-range as a 
function of $\Delta m$, defined as the
mass difference between the postulated non-standard Higgs and the
$\eta_b(1S)$ resonance \cite{mas03}. For the largest 
values of $\Delta m$
only the lower area of the shaded region would be likely acceptable
- corresponding to the highest estimates of  
${\cal P}^{\Upsilon}(\eta_b^*\gamma_s)$ in (\ref{eq:probability}). 

\begin{figure}
\epsfysize=5.5cm
\centerline{\hbox{{\hfil\epsfbox{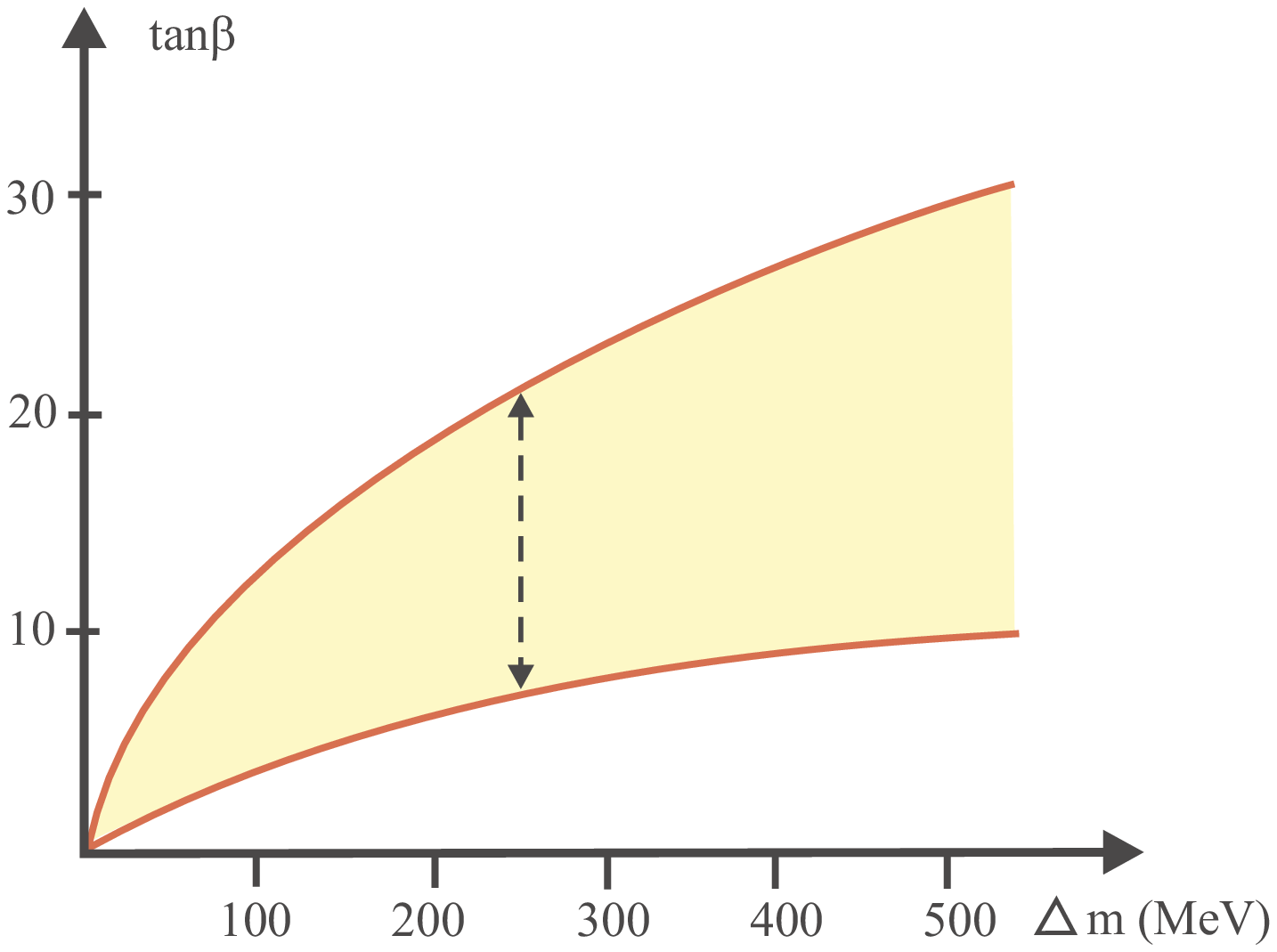}\hfil}
{\hfil\epsfbox{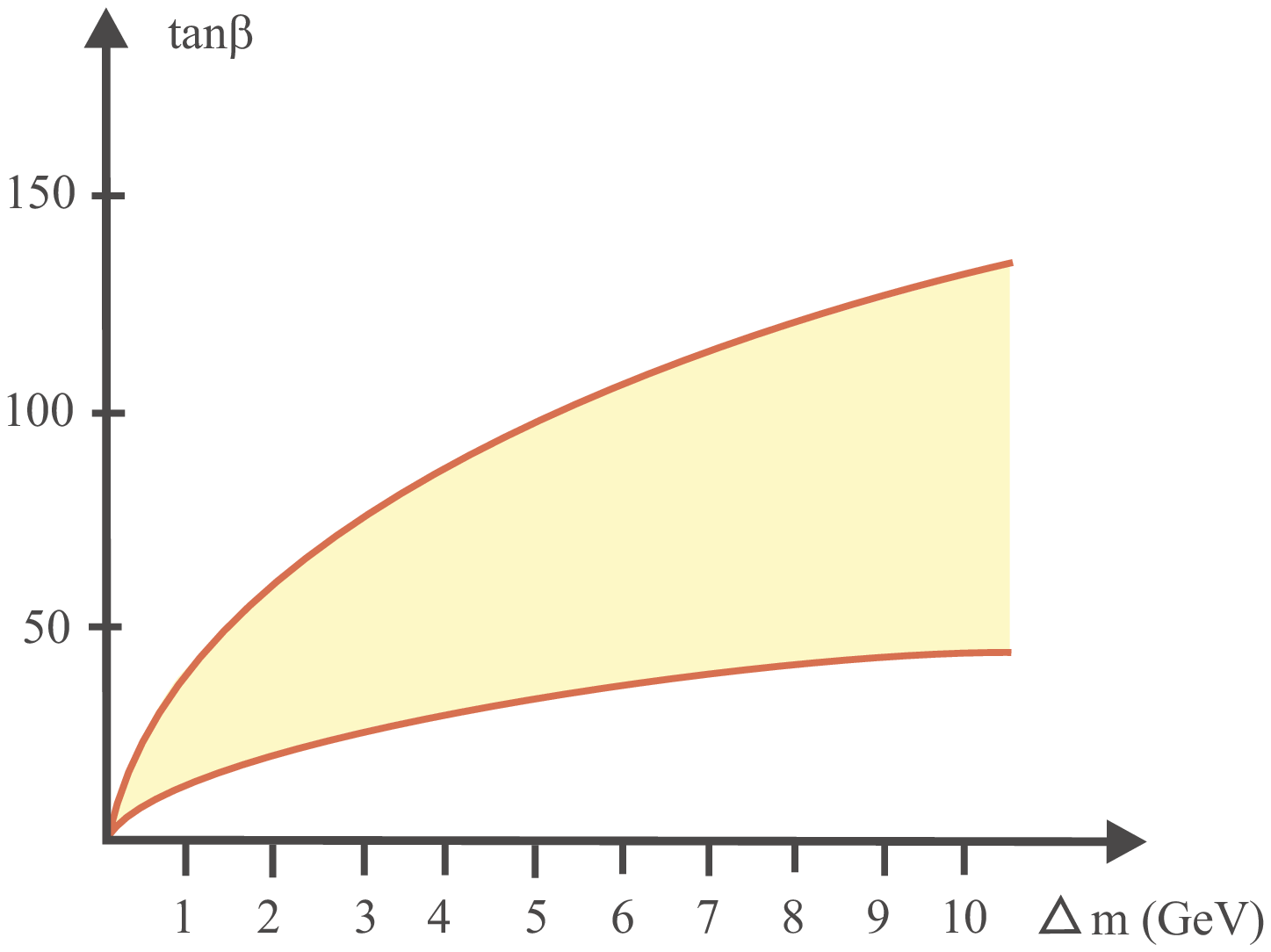}\hfil}}}
\caption{Required $\tan{\beta}$ values (shaded area) needed to
account for a $\sim 10\%$ breakdown of lepton universality in $\Upsilon$
decays as a function of $\Delta m$ according to
the factorization (\ref{eq:finalwidth}). The vertical dotted line
shows the range of $\tan{\beta}$ for
$\Delta m = 250$ MeV used in \cite{mas03} as a reference value.}
\end{figure}

On the other hand, applying second-order time perturbation theory
instead of the factorization given by Eq.(\ref{eq:factorization}), one obtains 
that the partial width of the whole decay, in the
narrow width approximation (i.e. $\Gamma_{\eta_b}/2 \ll \Delta E_{hs}$), 
reduces to 
\begin{equation}
\Gamma_{\,\Upsilon\to\gamma_s\,\ell\ell}\ \simeq\ 
\Gamma^{\,M1}_{\,\Upsilon\to\gamma_s\eta_b}
\frac{\Gamma_{\eta_b\to\ell\ell}}{\Gamma_{\eta_b}}
\label{eq:finalwidth2} 
\end{equation} 
In fact this equation matches a cascade decay 
taking place via the aforementioned
$\eta_b^*$ intermediate state. 
Let us remark that Eqs.(\ref{eq:finalwidth}) and (\ref{eq:finalwidth2})
are numerically (and conceptually) quite different as one expects
$\Gamma_{\Upsilon} \ll \Gamma_{\eta_b}$. Thus, in general, 
$\tan{\beta}$ values higher than in the previous approach
are required in order to account for the same ${\cal O}(10\%)$ 
breakdown of lepton universality, while the Higgs mass value
has to be kept below $B\bar{B}$ threshold to
enhance the decay rate through the Higgs propagator as already mentioned.

Now, since the Higgs-mediated partial width of the leptonic decay of
the $\eta_b$ state
depends on the fourth power of $\tan{\beta}$ 
and the leptonic mass squared (see
Eq.(20) of Ref.\cite{mas03}), a large value of the former quantity
would imply that the pseudoscalar resonance (which stands
below open bottom production)
would mainly decay through the tauonic channel, i.e.
\begin{equation}
\Gamma_{\eta_b}\ \simeq\ {\Gamma}_{\eta_b\to\tau\tau}
\label{eq:comparison2}
\end{equation}
overwhelming all other decay modes. For example, if 
$\tan{\beta}\ {\gtrsim}\ 35$ the full
width turns out to be $\Gamma_{\eta_b}\ {\gtrsim}\ 30$ MeV
in contrast with the expected 
$\Gamma_{\eta_b} \simeq 4$ MeV obtained from the
asymptotic expression $\Gamma_{\eta_b} 
\simeq m_b/m_c \times [\alpha_s(m_b)/\alpha_s(m_c)]^5 \times
\Gamma_{\eta_c}$, setting 
$\Gamma_{\eta_c(1S)}=16\pm 3\,$MeV \cite{pdg}
\footnote{Let us stress that there are still open windows
for relatively light non-standard Higgs masses and 
moderate $\tan{\beta}$ values, still not 
ruled out by direct searches
at LEP \cite{opal,opal2}. Also notice that for a 
meaningful existence of the pseudoscalar state, 
the $\eta_b$ full width should remain smaller than the 
$\Upsilon-\eta_b$ hyperfine splitting}. 

Then from Eqs. (\ref{eq:finalwidth2})
and (\ref{eq:comparison2}),
the partial width for the tauonic channel
approximately coincides with the M1-transition width, i.e. 
\begin{equation}
\Gamma_{\,\Upsilon\to\gamma_s\,\tau\tau}\ \simeq\ 
\Gamma^{\,M1}_{\,\Upsilon\to\gamma_s\eta_b} 
\label{eq:finalwidth3} 
\end{equation} 
because of the almost complete cancellation of the partial and full 
widths of the $\eta_b$. Next, dividing both sides of the above 
approximate equality by the $\Upsilon$ full width, one finds
\begin{equation}
{\cal B}_{\Upsilon\to\gamma_s\,\tau\tau}\ \simeq\ 
\frac{\Gamma^{M1}_{\Upsilon{\rightarrow}\gamma_s\eta_b}}
{\Gamma_{\Upsilon}}
\label{eq:finalwidth4} 
\end{equation}

As ${\cal B}_{\Upsilon \to \tau\tau} \sim 10^{-2}$ \cite{pdg},
the value of the BF's ratio ${\cal R}_{\tau}$ of 
Eq.(\ref{eq:ratio}), calculated
according to the above expression (\ref{eq:finalwidth4}), can indeed
reach a $10\%$ order-of-magnitude for the highest estimates 
from Eq.(\ref{eq:probability}). This result represents a
step forward in our investigation using this approach
with respect to Ref.\cite{mas03}.
Thus, a large $\eta_b$ width 
(e.g. $\Gamma_{\eta_b}\ {\gtrsim}\ 30$ MeV) could stem
either from (\ref{eq:finalwidth}), or 
from the alternative factorization (\ref{eq:finalwidth2}),
for high $\tan{\beta}$ in both cases.

Let us observe that quite broad pseudoscalar
$b\bar{b}$ resonances might explain why there is no evidence
found from hindered M1 radiative decays of higher Upsilon
resonances into $\eta_b(1S)$ and $\eta_b(2S)$ states
in the search performed by CLEO \cite{cleo02,cleo03}.
The corresponding signal peak (which should appear in the 
photon energy spectrum) could be considerably
smoothed - in addition to the spreading from the experimental
measurement - and thereby might not be significantly
distinguished from the background (arising
primarly from $\pi^0$'s). Of course, the 
matrix elements for hindered
transitions are expected to be small and difficult to 
predict as they are generated by relativistic and finite size
corrections. Nevertheless, most of the theoretical calculations
(see a compilation in Ref.\cite{godfrey})
are ruled out by CLEO results (at least) at $90\%$ CL,
though substancially lower rates are obtained in \cite{lahde} where
exchange currents play an essential role and therefore cannot be
currently excluded. 
Let us finally point out that a large full width of the $\eta_b$
resonance would bring negative effects on the prospects for its
detection at the Tevatron through the 
double-$J/\psi$ decay: $\eta_b \to J/\psi+J/\psi$. Indeed, the
expected BF would drop by about one order of magnitude with
respect to the range between $7 \times 10^{-5}$
and  $7 \times 10^{-3}$ assumed in \cite{braaten}.

Furthermore, another interesting possibility is linked to a
$A^0-\eta_b$ mixing \cite{drees} which could 
sizeably lower the mass of the mixed (physical) $\eta_b$ state, 
especially for high $\tan{\beta}$ values starting from similar 
masses of the unmixed states \cite{mas03}. Then  
the signal peak in the photon energy plot
could be (partially) shifted
off the search window used by CLEO \cite{cleo02,cleo03}
towards higher $\gamma$ energies (corresponding to a smaller $\eta_b$ mass
\footnote{This would be the case if the (unmixed)  
CP-odd Higgs boson had a mass greater than the (unmixed) 
$\eta_b$ resonance \cite{drees}. We work under this hypothesis throughout
this paper}) perhaps contributing additionally to the
failure to find evidence about the existence of the $\eta_b$
resonances to date. As a particular but illustrative example, assuming for 
the masses of the unmixed states $m_{\eta_{b0}} \simeq m_{A_0^0} =9.4$ GeV
and the moderate $\tan{\beta}=20$ value, one gets for the 
physical states $m_{A^0}= 9.56$ GeV and $m_{\eta_b}=9.24$ 
GeV respectively 
\footnote{The mass formula for the physical $A^0$ and $\eta_b$
states in terms of the unmixed
states (denoted as $A_0^0$ and $\eta_{b0}$ respectively),  
and the off-diagonal mass matrix element 
$\delta m^2 \simeq 0.146 \times \tan{\beta}$ \cite{mas03},
for quite narrow resonances
(i.e. $\Gamma_{\eta_{b0}},\ \Gamma_{A_0^0}\ \ll m_{\eta_{b0}},\ m_{A_0^0}$)
reads \cite{mas03}:
\[
m_{\eta_b,A^0}^2\ \simeq\ \frac{1}{2}(m_{A_0^0}^2
+m_{\eta_{b0}}^2)\ \mp\ \ \frac{1}{2}\biggl[(m_{A_0^0}^2
-m_{\eta_{b0}}^2)^2+4(\delta m^2)^2\biggr]^{1/2} 
\]
which yields in the case of the physical $\eta_b$ and $A^0$ particles
for different mass intervals:
\begin{eqnarray}
m_{\eta_b,A^0} & \simeq & 
m_{\eta_{b0}}\ \mp\ \frac{\delta m^2}{2m_{\eta_{b0}}} 
\ \ ;\ \ \ 0 < m_{A_0^0}^2-m_{\eta_{b0}}^2 << 2\ \delta m^2 \nonumber \\
m_{\eta_b,A^0} & \simeq & 
m_{\eta_{b0}}\ \mp\ \frac{(\delta m^2)^2}
{2m_{\eta_{b0}}(m_{A_0^0}^2-m_{\eta_{b0}}^2)} 
\ \ ;\ \ m_{A_0^0}^2-m_{\eta_{b0}}^2 >> 2\ \delta m^2 \nonumber
\end{eqnarray}}.

On the other hand, CLEO has completed detailed scans of the
$\Upsilon(nS)$ ($n=1,2,3$) resonances and we want to stress
the relevance of these measurements (aside many other applications)
for testing more accurately the possible existence
of new physics by a more precise determination of the 
electronic, muonic and tauonic BFs of
{\em all three} resonances below open bottom threshold. In case 
no lepton universality breaking is definitely found, some
windows in the $\tan{\beta}$-$m_{A^0}$ parametric space 
for such a non-standard CP-odd light Higgs boson 
\footnote{Likewise, other scenarios and models can be considered, e.g., 
Higgs bosons with no defined CP \cite{pilaftsis}}
would be closed. 
\par
\vskip 0.3cm
{\em Acknowledgements:} I thank R. Galik, H. Muramatsu,
and T. Skwarnicki for useful comments related to CLEO results.
I also thank J. Papavassilou and P. Ruiz for discussions
and critical reading of the manuscript.


\newpage

\thebibliography{References}
\bibitem{mas03} M.~A.~Sanchis-Lozano, 
arXiv:hep-ph/0307313.
\bibitem{mas02} M.~A.~Sanchis-Lozano,
Mod.\ Phys.\ Lett.\ A {\bf 17}, 2265 (2002)
[arXiv:hep-ph/0206156].
\bibitem{gunion} J.~Gunion et al., {\em The Higgs Hunter's Guide} 
(Addison-Wesley, 1990).
\bibitem{cleo02} A.~H.~Mahmood {\it et al.}  [CLEO Collaboration],
arXiv:hep-ex/0207057. 
\bibitem{cleo03} T.~E.~Coan, arXiv:hep-ex/0305045. 
\bibitem{pdg} K.~Hagiwara {\it et al.}  [Particle Data Group Collaboration],
Phys.\ Rev.\ D {\bf 66}, 010001 (2002).
\bibitem{bodwin} G.~T.~Bodwin, E.~Braaten and G.~P.~Lepage,
Phys.\ Rev.\ D {\bf 51}, 1125 (1995)
[Erratum-ibid.\ D {\bf 55}, 5853 (1997)]
[arXiv:hep-ph/9407339].
\bibitem{opal} G.~Abbiendi {\it et al.}  [OPAL Collaboration],
Eur.\ Phys.\ J.\ C {\bf 27}, 483 (2003)
[arXiv:hep-ex/0209068].
\bibitem{opal2} G.~Abbiendi {\it et al.}  [OPAL Collaboration],
Eur.\ Phys.\ J.\ C {\bf 23}, 397 (2002)
[arXiv:hep-ex/0111010].
\bibitem{godfrey} S.~Godfrey and J.~L.~Rosner,
Phys.\ Rev.\ D {\bf 64}, 074011 (2001)
[Erratum-ibid.\ D {\bf 65}, 039901 (2002)]
[arXiv:hep-ph/0104253].
\bibitem{lahde} T.~A.~Lahde, C.~J.~Nyfalt and D.~O.~Riska,
Nucl.\ Phys.\ A {\bf 645}, 587 (1999)
[arXiv:hep-ph/9808438].
\bibitem{braaten} E.~Braaten, S.~Fleming and A.~K.~Leibovich,
Phys.\ Rev.\ D {\bf 63}, 094006 (2001)
[arXiv:hep-ph/0008091].
\bibitem{drees} M.~Drees and K.~i.~Hikasa,
Phys.\ Rev.\ D {\bf 41}, 1547 (1990).
\bibitem{pilaftsis} M.~Carena, J.~R.~Ellis, S.~Mrenna, A.~Pilaftsis 
and C.~E.~M.~Wagner,
Nucl.\ Phys.\ B {\bf 659}, 145 (2003)
[arXiv:hep-ph/0211467].
\end{document}